%% file: main.tex
\documentclass{article}
\usepackage[final]{neurips_2025}
\usepackage{graphicx}
\usepackage{hyperref}
\usepackage{lmodern}
\makeatletter
\providecommand{\@trackname}{} 
\makeatother
\hypersetup{
    colorlinks=true,
    linkcolor=blue,
    urlcolor=blue,
    citecolor=blue
}

\title{DrP: Meta's Efficient Investigations Platform at Scale}

\author{
\textbf{Shubham Somani},
\textbf{Vanish Talwar}\thanks{Current affiliation: Moloco. Work conducted while at Meta.},
\textbf{Madhura Parikh},
\textbf{Eduardo Hernandez}, \\
\textbf{Jimmy Wang},
\textbf{Shreya Shah},
\textbf{Chinmay Gandhi},
\textbf{Sanjay Sundarajan}, \\
\textbf{Neeru Sharma},
\textbf{Srikanth Kamath},
\textbf{Nitin Gupta},
\textbf{Benjamin Renard}, \\
\textbf{Ohad Yahalom},
\textbf{Chris Davis}\thanks{Current affiliation: Roblox. Work conducted while at Meta.} \\
\\
Meta
}

\begin{document}
\maketitle
\begin{abstract}
\input{sections/0_abstract}
\end{abstract}

\section{Introduction}
\input{sections/1_introduction}

\section{Motivation \& Use cases}
\input{sections/2_Motivation_and_Usecases}

\section{DrP Overview}
\input{sections/3_Systems_Overview}

\section{DrP SDK}
\input{sections/4_SDK}

\section{DrP Backend}
\input{sections/5_Backend}

\section{Integrations \& Extensions}
\input{sections/6_Integrations_and_Extensions}

\section{Evaluation}
\input{sections/7_Top_level_experiments}

\section{User Studies}
\input{sections/7_1_User_Studies}

\section{Lessons Learned}
\input{sections/8_Lessons_learned}

\section{Related Work}
\input{sections/9_Related_Work}

\section{Conclusion \& Future Work}
\input{sections/10_Conclusion}

\section*{Acknowledgments}
\input{sections/Acknowledgements}


\nocite{*}
\bibliographystyle{abbrv}

\end{document}

%% file: sections/0_abstract.tex
Investigations are a significant step in the operational workflows for large scale systems across multiple domains such as services, data, AI/ML, mobile. Investigation processes followed by on-call engineers are often manual or rely on ad-hoc scripts. This leads to inefficient investigations resulting in increased time to mitigate and isolate failures/SLO violations. It also contributes to on-call toil and poor productivity leading to multiple hours/days spent in triaging/debugging incidents. In this paper, we present DrP, an end-to-end framework and system to automate investigations that reduces the mean time to resolve incidents (MTTR) and reduces on-call toil. DrP consists of an expressive and flexible SDK to author investigation playbooks in code (called analyzers), a scalable backend system to execute these automated playbooks, plug-ins to integrate playbooks into mainstream workflows such as alerts and incident management tools, and a post-processing system to take actions on investigations including mitigation steps. \\
We have implemented and deployed DrP at large scale at Meta covering 300+ teams, 2000+ analyzers, across a large set of use cases across domains such as services, core infrastructure, AI/ML, hardware, mobile. DrP has been running in production for the past 5 years and executes 50K automated analyses per day. Overall, our results and experience show that DrP has been able to reduce average MTTR by 20 percent at large scale (with over 80 percent for some teams) and has significantly improved on-call productivity.

%% file: sections/1_introduction.tex
Large scale and globally deployed services in private/public clouds are prevalent in multiple industry verticals today including social media, e-commerce, gaming, financial, transportation and more. These services rely on application specific micro-services and infrastructure such as data warehouses, ML/AI systems, data analytics, core platforms. End to end operations is a key functionality needed to ensure the SLOs for these services and infrastructure such as availability, reliability, and performance. Typical operational workflows can be broken down into three main components (a) alerting for quick detection of incidents such as outages or failures, (b) investigations to triage and mitigate the incidents, and (c) deep-dive root cause analysis for long term fixes of the incident triggers. While there has been a large body of work to improve alerting \cite{chandola_alerting} and similarly a large number of domain specific root cause analysis techniques \cite{lou2020understanding, kim2013root, hoffmann2018snailtrail, teoh2019perfdebug, shao2019griffon}, the middle step of end-to-end investigations to quickly triage the incident to isolate and mitigate the problem has largely been ignored. With the advent of large scale systems that consist of multiple domains/components and have complex dependencies, this step is quickly becoming critical to have smooth and scalable operations.

\par
\begin{figure}[h!]
    \centering
    \includegraphics[width=0.8\linewidth]{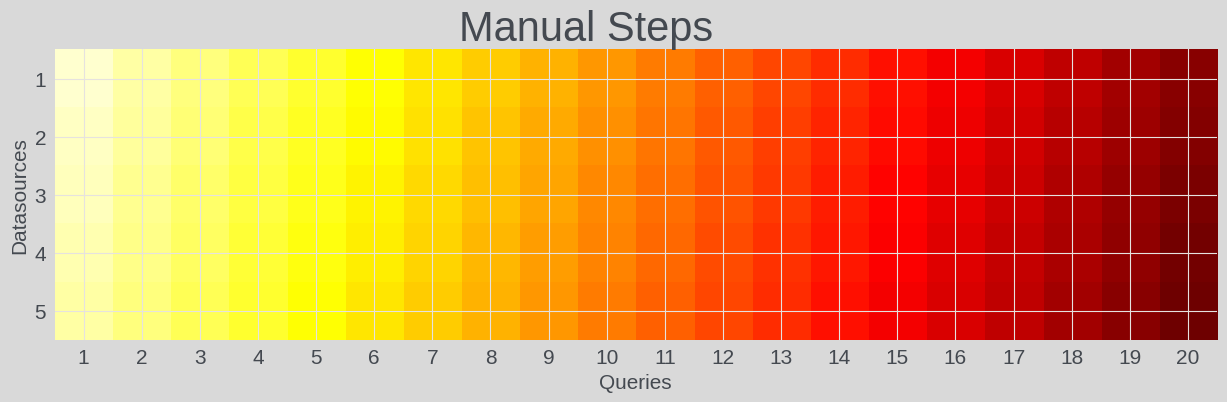}
    \caption{Rapid increase in number of investigation steps with increasing complexity. Complex systems typically require a lot more queries over multiple data sources to triage the incident.}
    \label{fig:intro-steps}
\end{figure}

 For example, let’s say we have a globally deployed service and an outage happens. The outage could be from a particular region and/or could be due to client, network, hardware, or other dependency issues. To investigate, we have to look at multiple sources of data and navigate through a complex ecosystem of tools and datasets to isolate the problem. Current approaches to do such investigations are largely manual and rely on outdated playbooks in wiki pages. Figure \ref{fig:intro-steps} shows how quickly these steps can increase as complexity of systems increases. Moreover, these steps are repetitive. On-calls \footnote{On-calls are engineers designated to look into operational issues, typically on a rotation \cite{oncall}} repeat the same steps for multiple incidents and the manual process is error prone. Long investigations involving multiple engineering teams are quite typical leading to prolonged downtimes and on-call toil.

\par We address such inefficient investigations in this paper focusing on a general framework to build automated systems and techniques for investigations that can be applied to a wide range of end-to-end application and infrastructure use cases. Our goal is to improve two important measures - mean time to resolution (MTTR) and on-call toil. Mean time to resolution is the average time to triage and mitigate incidents before the system is back and healthy. On-call toil is the amount of hours on-calls spend triaging incidents. The two measures are related but target different goals - system availability and on-call productivity.

Our overall approach is to {\em codify} the manual playbooks and tribal knowledge prevalent in organizations. We provide an expressive and flexible SDK to author investigation playbooks (in Python, PHP) including a rich set of helper libraries for data access and problem isolation analysis (eg. timeseries correlation, event correlation, dimension analysis). We call such codified playbooks {\em analyzers}. To execute these analyzers, we provide a scalable backend providing both multi-tenant and isolated analyzer execution. To consume the analyzers, we provide comprehensive integrations with alerting and incident management tools. This allows for auto-trigger of analyzers on incidents. For example, when an alert fires, the appropriate analyzer is automatically executed and the results of the analysis are included on the alert page or task. We also provide for ad-hoc execution of analyzers and have built easy-to-use UI and CLI interfaces that are used by several hundreds of engineers every day at Meta. Furthermore, we provide a post-processing system that allows for automated actions to take based on the analysis results. For example, if the alert is due to wrong config changes, the post processing system can create tasks and PRs to mitigate the issue.

Building a dedicated system for automated investigations provided interesting new challenges and opportunities. {\em First}, many of the services in large organizations depend on other services. An analyzer trying to triage an issue with a service has to look into whether any of the dependent services has issues or not. By codifying playbooks, this now just requires a call into the analyzer of the dependent service. The dependent service may depend on other services and so on, thus leading to a chain of analyzer calls. We provide first class support for such chaining in our SDK. {\em Secondly}, building a general runtime system for executing analyzers required us to provide several interesting system hacks. Since analyzers are constantly in churn (due to playbook updates) and with a constant growth of analyzers, we had to build a runtime system that can dynamically import new analyzers without requiring any restart or downtime of the runtime system. We also built a customized CI/CD\footnote{Continuous Integration/Continuous Deployment} system to allow for analyzer package updates. {\em Thirdly}, providing a testing framework for analyzers is extremely hard since there is usually no good way to record past incidents. Building elaborate unit tests is also infeasible given the many different paths that an investigation can take. We created a novel backtesting framework for the analyzers that reuses incidents from previous runs of analyzers to test updates to the analyzer. This helped to catch bugs or problems earlier in the development cycle at code PR (Pull Request) time improving analyzer quality. Our CI/CD system also canaries any changes to analyzers before deploying them to production. This has prevented the deployment of buggy analyzers, significantly improving the reliability and availability of our backend system.

In summary, we make the following contributions:

\begin{enumerate}
    \item[\textbullet] We present DrP, a first known general purpose framework for automated playbook execution for efficient investigations at scale. DrP consists of an expressive, flexible SDK for creating automated playbooks called {\em analyzers}, and a scalable backend system to execute the analyzers. 

    \item[\textbullet] We describe various integrations for DrP with alerting and incident management tools at Meta simplifying the investigation processes at large scale. We also describe extensions to allow post-processing actions to be taken based on the automated investigations.

    \item[\textbullet] We present production experience with DrP from its usage for a large number of use cases at Meta for the past 5 years. DrP has a total of 2000+ analyzers used by 300+ teams running over 50K automated analysis per day. DrP has resulted in significant MTTR savings for teams, with some over 80\%, and on average 20\% improvement across multiple use cases at scale. DrP has also received positive feedback in several surveys validating reduction in on-call toil.
\end{enumerate}

%% file: sections/2_Motivation_and_Usecases.tex
\label{section:motivation}

Managing large-scale distributed systems across thousands of services and servers in global data centers is inherently complex. Our focus is on the challenge of investigations, which involves triaging alerts to pinpoint the source of issues for effective incident mitigation. As system complexity and telemetry data volume increase, investigations become increasingly daunting. Below, we present examples of investigation scenarios and summarize the associated challenges.

\subsection{Investigation Examples}

\label{investigation_examples}

\noindent {\bf Services Debugging:} Modern systems rely on interconnected services with dependencies. In large-scale deployments, these services are globally distributed with multiple instances for load balancing and availability. A service discovery and routing system directs RPC calls to the appropriate service instance. Operations set up alerts for service SLOs related to errors, latency, and throughput. When an alert indicates an increased error rate, the on-call team must investigate various dimensions to isolate the issue. For instance, are errors concentrated in a specific region or cluster? Are they limited to certain service endpoints or affecting all API calls? Is a dependent service causing the problem? The issue could stem from network, hardware, client, or service bugs. Isolating the problem requires querying multiple data sources, performing dimension analysis, and reviewing related alerts, making it a time-consuming and complex exercise.

\begin{figure}[h!]
    \centering
    \includegraphics[width=1.05\linewidth]{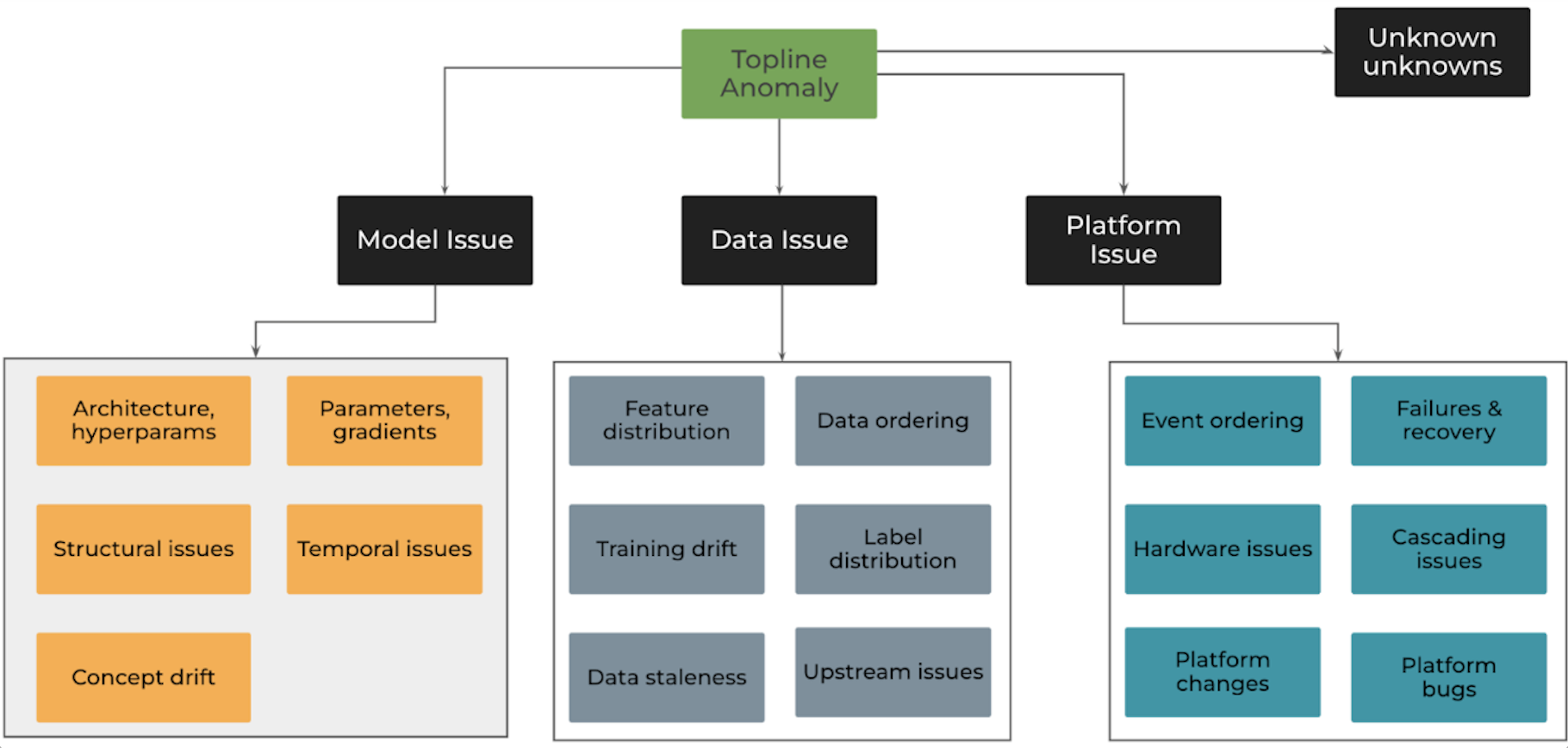}
    \caption{Example of a complex investigation - ML prediction debugging}
    \label{fig:ml_debugging}
\end{figure}

\noindent {\bf ML Debugging:} Debugging ML systems is highly complex. When an alert triggers for ML prediction metrics, the on-call team must navigate a complex investigation tree. Is the anomaly due to a model issue from training or architecture changes? Is it a feature issue during inference or training, possibly linked to upstream data sources? Or is it a platform issue requiring low-level observability metrics inspection? These scenarios involve numerous steps and complexities in AI/ML issue investigations as shown in \ref{fig:ml_debugging}.

\noindent {\bf App Debugging:} Debugging mobile apps is challenging. Increased bug reports from a mobile app prompt several investigation steps for the on-call team. Is the issue related to a recent update, backend service changes, specific OSes or device models, certain events, or third-party libraries? Mobile devices' limited resources complicate detailed log collection and debugging, making problem isolation more difficult.

\subsection{Summary of Investigation Challenges}

Here are common challenges faced in large-scale investigations:

\par \noindent {\bf Time-consuming manual steps:} Current investigations are often manual and repetitive, requiring on-call engineers to sift through multiple dashboards and query various data sources. This process can be time-consuming, delaying incident mitigation and consuming significant engineering hours.

\par \noindent {\bf Outdated or incomplete playbooks:} On-call engineers rely on playbooks in wikis for alert triage. These wikis often become outdated or incomplete, forcing engineers to depend on tribal knowledge and ad-hoc solutions. This results in unsynchronized scripts or steps, hindering knowledge sharing.

\par \noindent {\bf Bottlenecks caused by too few experts:} Debugging expertise is often limited to a few subject-matter experts (SMEs), creating bottlenecks during investigations. This lack of knowledge transfer makes it difficult to distribute expertise across the team.

\par \noindent {\bf On-call toil and burnout:} Engineers may face numerous incidents during rotations. Manual processes are not scalable, leading to many incidents being unresolved. Repeated investigations consume bandwidth, preventing root cause resolution.

\par \noindent {\bf Inadequate support for complex systems:} Teams may use scripts, dashboards, and notebooks for investigations, but these often fail to support complex systems with multiple domains. Maintaining these tools is challenging, and there is no standardized process for their upkeep.

%% file: sections/3_Systems_Overview.tex
In this section, we introduce DrP, our solution to the investigation challenges discussed earlier. Our approach involves creating automated debugging playbooks by transforming wiki playbooks, ad-hoc scripts, and tribal knowledge into code. DrP provides an end-to-end investigations framework to author these playbooks, execute them at scale, and integrate them with operational workflows. The goal is to reduce MTTR and reduce oncall toil by automating repeated investigations.

Developing such an automated investigations framework requires us to address several open questions - 
\begin{itemize}
\item How to make it easy for developers to author and test the automated playbooks?
\item How to ensure high quality automated analysis to build confidence and trust for use in production incidents?
\item How to generalize the framework to cater to diverse use cases and integrate it with org-wide operational workflows?
\item How to scale the execution backend to cater to hundreds of thousands of automated analysis?
\end{itemize}

\begin{figure}[h!]
    \centering
    \includegraphics[width=1.05\linewidth]{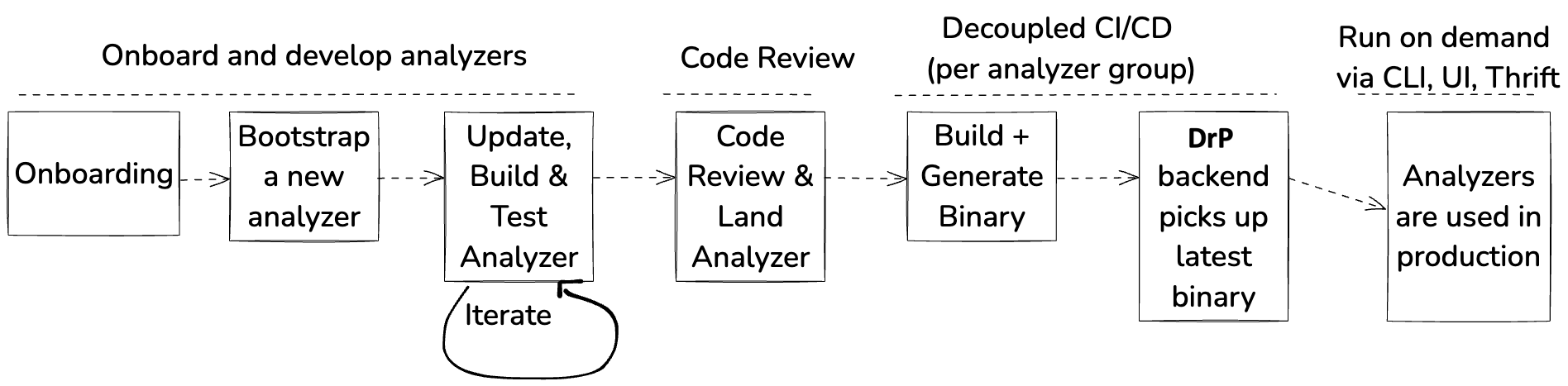}
    \caption{High-level Authoring Overview}
    \label{fig:authoring-overview}
\end{figure}

DrP provides an SDK to author the automated playbooks, called {\em analyzers}. The SDK offers catered APIs and libraries to make it easy for developers to codify their investigation workflows. Typically, teams would first enumerate their investigation steps, then utilize the DrP SDK and bootstrap utilies to create a template analyzer, thereafter they would iteratively code in their investigation decision tree into the analyzer levearging the SDK's data access and analysis libraries. To perfect the investigation workflow, DrP developers often undergo multiple "code, build, and render" cycles. To reduce friction for developers and to boost productivity, DrP also provides a WYSIWYG \footnote{What You See Is What You Get} authoring experience via a Visual Studio Code extension. This allows developers to save code changes in the IDE \footnote{Integrated Developement Environment} and instantly see results in a UI interface, making the authoring and testing process intuitive and seamless. Figure \ref{fig:authoring-overview} shows the end-to-end authoring workflow with DrP.

To ensure analyzers are high quality, DrP provides access to a wide variety of analysis libraries, that can provide either rule based, statistical, or machine learning based techniques. Users can also bring in their own custom analysis libraries and plug in to their analyzers. DrP also provides testing utilities for unit tests, backtesting against past incidents, and canary testing. This adds to the quality of the analyzers and their reliable execution in production.

Once the analyzers are tested and code reviewed, they are packaged into executable binaries that can be used in multiple ways. In production, analyzers integrate with the Meta's operational workflows such as dashboards, CLIs, alerts, and incident management systems. Dedicated portals can also invoke any DrP analyzer with necessary inputs, and DrP UI widgets can be added to dashboards. Analyzers can automatically trigger upon alert activation, providing immediate results to on-call engineers and improving response times.

Analyzer invocation calls a scalable DrP backend which manages a queue for requests and a worker pool for secure execution. Results return asynchronously, with timeouts to prevent resource exhaustion. Comprehensive monitoring and logging support debugging. A separate post-processing system handles analysis results, annotating alerts with results of the automated analysis. The DrP Insights system periodically analyzes outputs to identify and rank top alert causes, aiding teams in prioritizing reliability improvements.

Overall, the combination of the flexible SDK, analysis libraries, testing, integrations, and scalable backend helps DrP address the challenges and open questions in building a generic one-size-fits-all investigations framework catering to a diverse set of use cases. In the next few sections, we describe some of DrP's components in more detail.

%% file: sections/4_SDK.tex
This section provides details on DrP SDK developed to streamline the authoring of DrP analyzers and post processing tasks. Our goal is to provide a user friendly and comprehensive SDK to allow users to quickly codify their investigation steps. 

\subsection{Authoring APIs} 

DrP SDK provides several useful authoring APIs, classes, and libraries in Python and PHP, enabling users to write analyzers in their preferred language and utilize domain-specific libraries. Some of the key ones are described below.

Any investigation requires appropriate context, for example, incident details such as alert ID, service name, threshold violations, and locations of telemetry data such as metrics, logs, traces. DrP SDK provides a {\em Context} class, a key-value dictionary for storing these investigation related parameters. A set of input APIs are provided to capture these inputs from various sources and are stored in the {\em Context} class after validating them for type and correctness. Any custom inferred inputs can also be dynamically added to the {\em Context} class throughout the lifetime of the analyzer execution.

\begin{figure}[h!]
    \centering
    \includegraphics[width=1.2\linewidth]{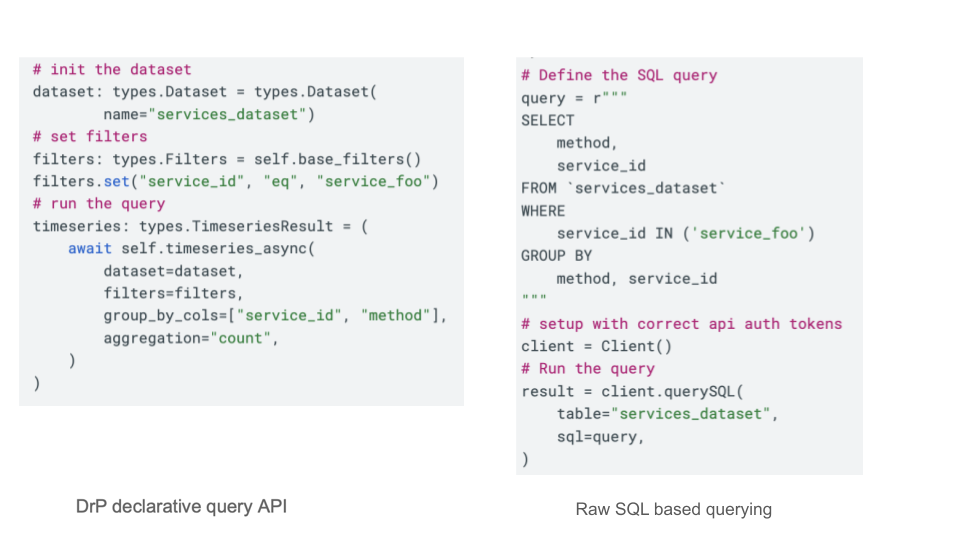}
    \caption{DrP declarative query API vs Raw SQL based querying}
    \label{fig:sdk-code-snippet}
\end{figure}

DrP also provides declarative, strongly typed APIs to query various data sources for telemetry data, e.g., time series databases, analytical databases, data warehouses, log databases. These APIs are tailored to common investigation query patterns. This has two key benefits: (1) eliminating the need for hard-coded SQL queries, which are difficult to maintain and debug; and (2) enabling easy query reuse across analyzers. Figure \ref{fig:sdk-code-snippet} illustrates DrP' Python API, showcasing its simplicity compared to raw SQL strings. Similar APIs exist for all supported data source categories, complete with error checking and logging.

To analyze the telemetry data, DrP provides convenient classes and libraries for data examination and issue isolation. The output from the data access APIs mentioned above are conveniently passed to the analysis libraries. DrP SDK also provides APIs for analyzer chaining to invoke other dependent analyzers to isolate dependency issues that may have caused the incident. More details about analysis libraries and analyzer chaining are explained later in this section.

After the analysis is complete, the output of the investigation is captured in a robust and structured \texttt{Findings} class that is essential for quick on-call use and system integration (alerts, incident management, UI, CLI, post-processors). The Findings class supports flexible rendering and evidence inclusion, offering outputs in plain text or machine-readable formats, including Thrift\cite{thrift}  payloads with self-describing schemas. These structures facilitate metadata addition for standardized UI widgets and custom React components for unique visualizations, enabling downstream processing and analytics.

Finally, the DrP SDK offers APIs for creating custom post-processors, which can be linked to any analyzer. The post-processing APIs and libraries simplify extracting analysis outputs from analyzers for custom actions, with boilerplate code for common tasks such as creating tasks, annotating alerts, and generating PRs. Analyzer findings are structured for parsing and dynamic post-processing at runtime, enabling advanced auto-triage or auto-mitigation actions. These post-processors are executed by a dedicated postprocessing tier for automated actions based on investigation outputs. 

\subsection{Analyzer Chaining}
DrP SDK introduces analyzer chaining, enabling users to execute analyzers in a specific sequence or as a DAG (Directed Acyclic Graph). This feature is popular due to the commonality of services having dependencies, including applications, storage, infrastructure, and hardware. Analyzer chaining allows for analyzers to call into other analyzers, for e.g., analyzers of dependent services, forming a chain or DAG for complex investigations.

Key functionalities for analyzer chaining include: (1) passing inputs and context to dependent analyzers, allowing temporary context overrides for additional parameters; (2) flexible outputs via the Findings class, enabling dependent analyzers to provide text and machine-readable outputs, which the calling analyzer can parse for relevant information; (3) lazy import of analyzers, allowing dynamic chaining without upfront latency.

\begin{figure}[h!]
    \centering
    \includegraphics[width=0.7\linewidth]{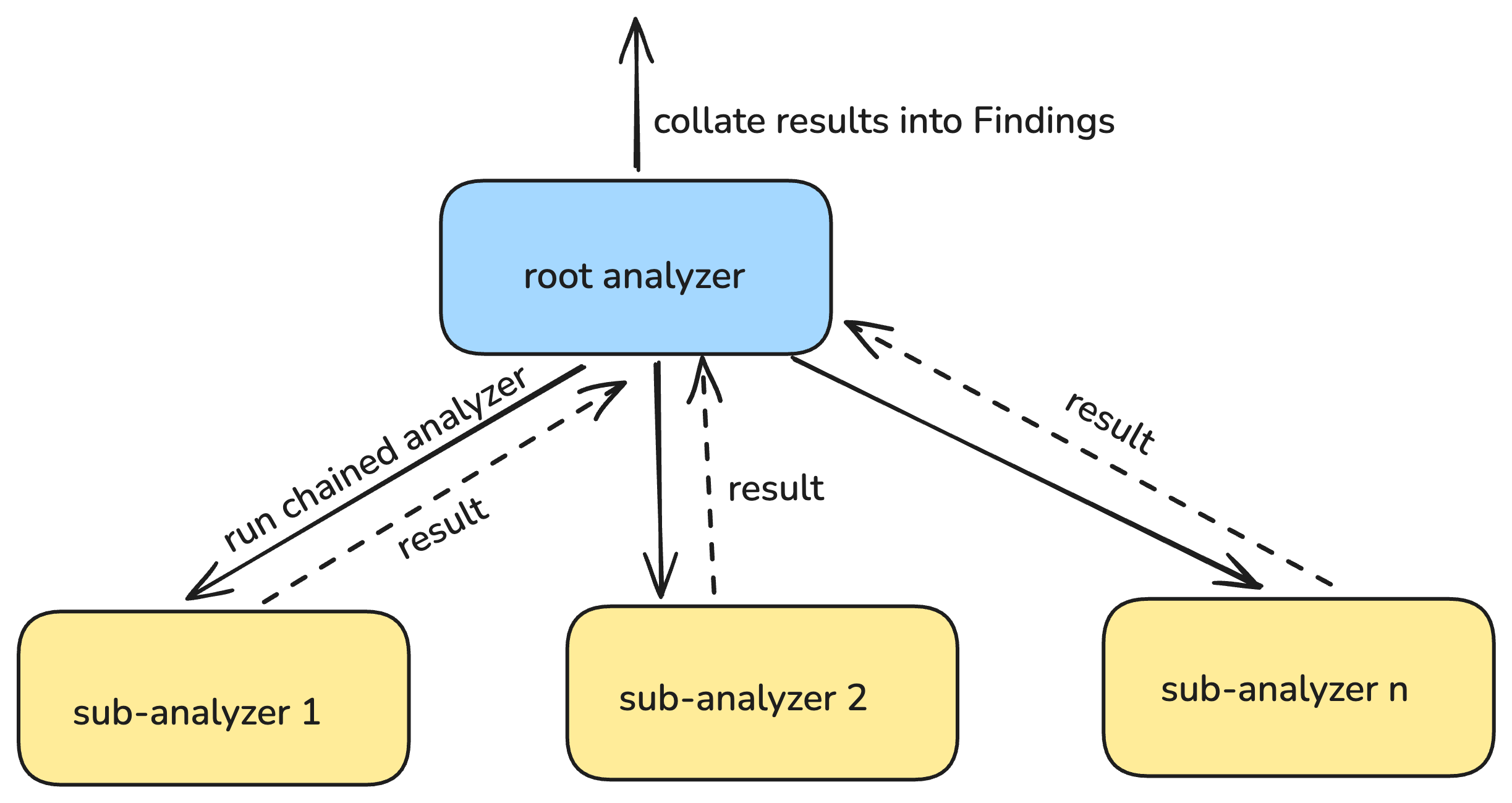}
    \caption{Example of Analyzer Chaining}
    \label{fig:analyzer-chaining}
\end{figure}

This feature promotes analyzer reuse, enabling domain experts to code workflows once for others to leverage. Cross-platform support allows chaining between PHP and Python analyzers, enhancing investigation capabilities. Figure \ref{fig:analyzer-chaining} illustrates a simple scenario where a top-level analyzer delegates to sub-analyzers and compiles their findings into a comprehensive result.

\subsection{Analysis libraries} 
Services in large scale systems generate large amounts of observability data. To tackle this problem, DrP has a library of scalable and efficient analysis algorithms based on statistical and ML techniques, including fast dimensional analysis \cite{lin2020fast}, anomaly detection\cite{toledano2018real}, event isolation\cite{luo2014correlating}, and time series correlation. A major challenge with these algorithms is leveraging them in real-time, latency-sensitive investigations. To speed-up investigations, various optimizations are available in the SDK based on common requirements. For example, for the most frequent repeated investigations, there is a pre-aggregation layer that can be applied on top of existing data that can reduce dataset size by up to 500X and significantly speed up analysis.

A common cause of incidents in large scale systems is attributed to code changes and deployment of config changes. DrP includes automated event isolation libraries to help accelerate investigations for such cases. Event isolation assistance has a suite of ML-based ranking models. These ML models leverage signals such as text matching, time correlation with alerts, and context around on-call to rank events with high and medium confidence. The models ranks thousands of code and config change events to isolate the issue. On average, this approach filters out majority of the uninteresting events during an active investigation; it ranks the  most probable suspicious events for the on-call engineer. The libraries provides annotations explaining the ranking and confidence, making it transparent for engineers.

In addition, DrP also allows users to bring in their own customized analysis library. This approach has helped scale the library and support diverse use cases. This provides a flexible and community based approach to the DrP SDK.

\subsection{Backtesting}

The DrP SDK provides unit test libraries with mock classes for analyzer testing. However, due to dynamic investigations and frequent updates, unit tests often miss coverage. Initially, manual integration tests via CLI were recommended, but they were hard to enforce, causing runtime failures.

To solve this, we developed a novel backtesting mechanism for analyzers. We retain inputs and outputs from past analyses, enabling integration tests on historical data (default 30 days) for modified analyzers. These tests filter out non-logic errors, highlighting code change issues. Automated in the PR review process, they block PRs until errors are fixed. Machine resources are dynamically allocated from a reserved pool for backtesting.

Backtesting has greatly improved analyzer quality and reduced failures. DrP also supports canary testing, detailed in Section \ref{section:backend}.

%% file: sections/5_Backend.tex
\label{section:backend}

This section details the DrP backend, which executes analyzers and post-processing tasks discussed earlier. Our aim is to develop a scalable, secure backend that ensures isolation among analyses and integrates seamlessly across various consumption surfaces.

Figure \ref{fig:backend} shows the high level overview of the backend. It provides both sync and async APIs to execute analyzers. The life-cycle for an asynchronous non-blocking request to DrP backend is as follows:

\begin{figure}[h!]
    \centering
    \includegraphics[width=1.0\linewidth]{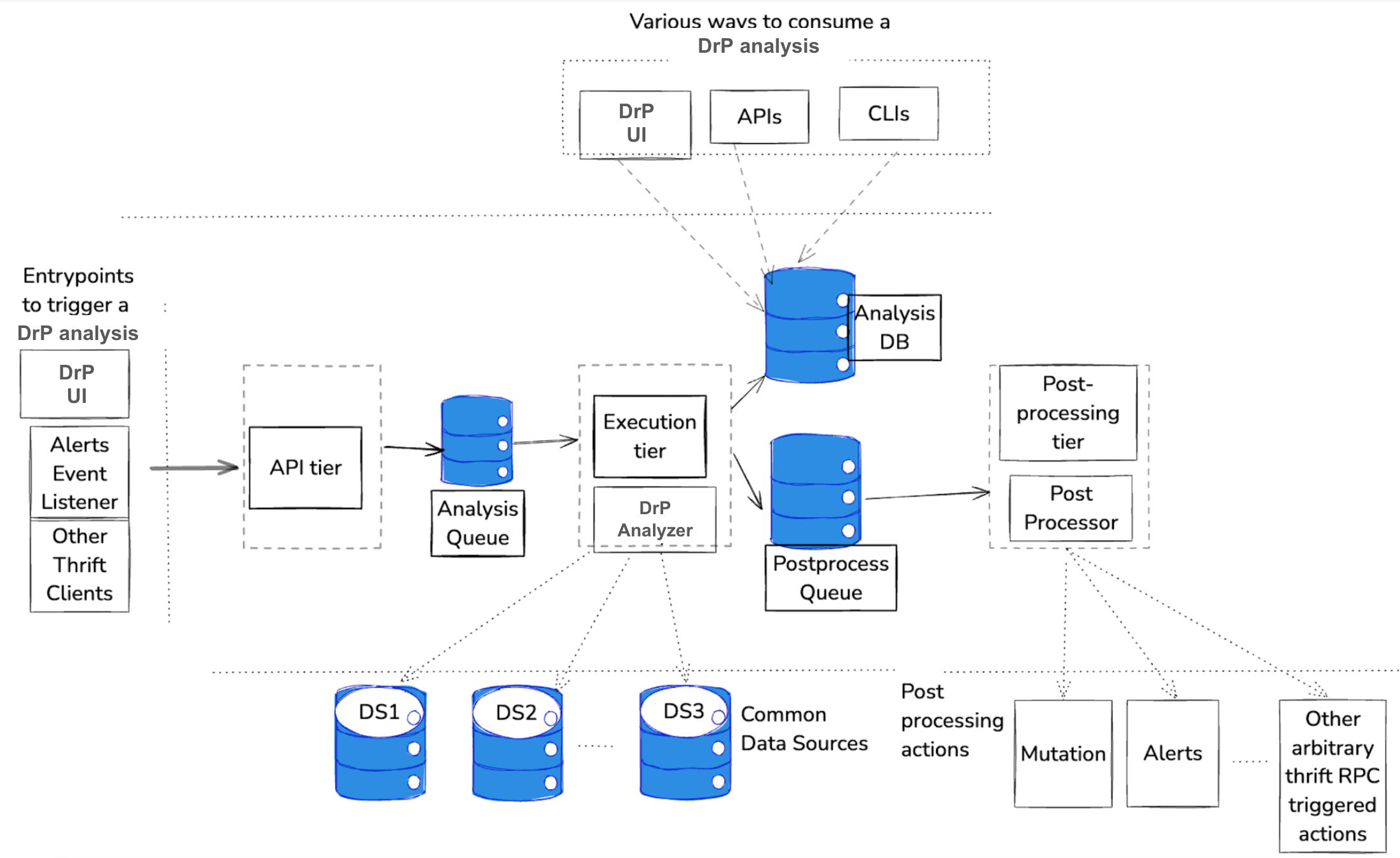}
    \caption{High level backend overview}
    \label{fig:backend}
\end{figure}

\begin{enumerate}
    \item User or automation sends the request to DrP's backend endpoint. The request includes details on which analyzer to run and the context of the incident to investigate.
    \item DrP's backend endpoint sends the request to a Request Queue to execute the request asynchronously.
    \item Worker tiers periodically dequeue requests from the queue service to process. The request is eventually picked up by a worker and reserved for execution.
    \item The worker tier starts an executor to process the request running the analyzer with the provided context as a CLI in a sandbox environment.
    \item Users call the peekDiagnoseStatus endpoint on the DrP backend to check the progress of the analyzer. If the request is still queued or the analyzer is running, users can check again later. If a user checks after the analyzer is finished running, the endpoint will either return the analysis findings, or the error if the analyzer did not succeed.
\end{enumerate}

\subsection{Scalable Execution}
DrP relies on a MySQL backed queue store to persist the requests. Each entry in the queue contains the following fields:
Request ID, Request timestamp, Analyzer identifier, Analysis context, Request status (Queued, Running, Success, Failed)

The worker tier operates on a reserved pool of machines, each running a runtime system, called Executor, that regularly queries request queues. The Executor parses requests and runs the specified DrP analyzer. Analyzers on an allowed list are run as server libraries, while others run as subprocesses. For PHP analyzers, the Executor uses RPC to communicate with another dedicated worker tier. The Executor monitors execution, terminating analyzers that exceed timeouts, and retrying failed executions up to a set limit. Once execution completes, times out, or reaches the retry limit, the request is removed from the queue. Worker tiers are monitored for utilization and queue times to adjust capacity as needed.

As mentioned above, the Executor is responsible for running the requested analyzer. Since we have a lot of analyzers (2000+) and growing, packaging all the analyzers into a single binary increases the size of the binary and it also takes a long time to load all the analyzer classes at runtime. It also leads to noisy neighbor issues when failures in one analyzer leads to failing of the whole binary. Instead we create smaller {\em analyzer groups} which are created based on expected affinity of analyzers. Each analyzer group has its own binary. 

When a request comes for an analyzer, we identify the analyzer group it belongs to and dynamically fetch the binary for that analyzer group from our package management system. Once fetched, we launch the binary as a separate sub-process to execute the requested analyzer. Since all dependent analyzers are packaged into one binary the chaining process is seamless in the analyzer group binary.

Dynamically fetching analyzer group binaries is a tradeoff we made between pre-loading all the analyzer groups at backend service start which can create extremely long delays in service start time versus {\em lazy loading} them at analyzer execution time with relative shorter delays. Lazy loading can however also lead to undesirable delays in analyzer execution which can lead to poor user experience. We did a study and found that 85 percent of the backend traffic comes from 10 percent of the top used analyzers. To ensure good user experience for those top analyzers, we pre-load the analyzer group binaries for those analyzers during the backend service startup. The rest are lazy loaded at the time of analyzer execution. This has provided us with a good balance of reasonable quick backend service startup and acceptable delays for dynamic fetching of analyzer group binaries. For certain very frequently used binaries, we have also gone a setup ahead and added those analyzers as libraries within the Executor binary itself. This causes negligible overhead in running those analyzers.

The DrP Runtime class manages analyzer execution, tracking context and dynamic variables. After execution, it persists analysis results in an MySQL-backed storage, including input details. Errors and debugging information are logged to an analytical database, allowing for historical result tracking and alerting analyzer owners on execution failures.

\subsection{Builds and Continuous Deployment}

After merging a PR for an analyzer, a continuous build process packages it into a production binary. Each analyzer is assigned to an analyzer group during authoring, with separate build processes for each group. A multi-step CI/CD system deploys these builds, including integration and canary testing. During canary testing, a sample of production traffic is run on the canary tier. Deployment is halted if errors occur. If tests pass, the latest changes are committed, and the analyzer is packaged into the group binary.


\subsection{Post processing Tier}
DrP features a post-processing tier for executing automated actions based on analyzer results, such as notifications or database updates. This involves a Post-processor QueueStore and a post-processing tier. If specified in the request, the Executor on analyzer worker tiers creates a PostProcessRequest, adding it to the Post-processor  QueueStore. The post-processing tier's runtime system dequeues and executes these requests. Results are stored post-execution, and the request is dequeued. A key challenge is idempotency, as post-processors may alter states like databases. We distinguish between stateful and stateless post-processors; stateful ones are not retried on failure, while stateless ones can be retried a configurable number of times.

%% file: sections/6_Integrations_and_Extensions.tex
DrP offers various integration options to meet diverse use cases, facilitating seamless analyzer execution and adoption. Below is a summary of key integrations:

\par \noindent {\bf Alerts Integration:} DrP is integrated with Meta's alerting systems, allowing users to specify analyzers for execution upon anomaly detection. When an alert triggers, the integration extracts metadata and creates input parameters for the analyzer. An RPC call to the DrP backend executes the analyzer, and post-processing generates a notification with findings on the alert page and task. Customized DrP widgets in the Alerts manager display findings user-friendly. The process from alert to analysis output typically takes a few minutes, providing near real-time analysis for on-call teams. This integration is widely used, significantly enhancing the on-call experience by enabling quick issue identification.

\par \noindent {\bf Dashboards:} DrP integrates with existing team dashboards through a UI widget, facilitating quick adoption without altering workflows. Many teams utilize this approach. Additionally, a standalone DrP UI is available for direct analyzer execution. Both the widget and standalone UI interface with the DrP backend for analyzer execution.

\par \noindent {\bf Service Level Objective Tool Integration:} Meta uses a tool for specifying and monitoring service level objectives (SLOs). When SLOs are violated, DrP automatically triggers an analyzer to investigate the cause. The analysis occurs in the DrP backend, and results in an auto-annotation of the SLO alert with the cause, selected from predefined categories, eg. NETWORK, CLIENT, SERVER. These annotations are stored and aggregated daily, with results displayed on the SLO tool dashboard. This helps teams identify and prioritize reliability projects based on top SLO violation causes. DrP's structured, automated investigations enable systematic SLO improvements, offering a unique capability at scale.

\par \noindent {\bf Parallel DAG Execution:} We have also integrated DrP with our in-house workflow management system to run analyzer DAGs (Directed Acyclic Graphs) that are long running and need parallelization. This allows scheduling and executing the DAG in a distributed cluster, enabling parallelism and preventing timeouts. 

\par \noindent {\bf Command Line Interface:} DrP offers a standalone CLI for executing analyzers with input parameters. Analyzers run remotely in the DrP backend, and results are displayed in text with flexible formatting options for usability.

DrP is also integrated with other incident investigation tools at Meta. We offer DrP as a library, allowing custom integrations and flexible analyzer execution.

%% file: sections/7_Top_Level_Experiments.tex
In this section, we present evaluation studies of DrP based on production experience at Meta. We focus on system adoption, backend scalability, MTTR improvements, developer and on-call experience. In Section \ref{case_studies}, we also present case studies from few teams.

\input{sections/7_Top_level_experiments/7_2_System_Adoption}
\input{sections/7_Top_level_experiments/7_1_System_Performance}
\input{sections/7_Top_level_experiments/7_4_Developer_Experience}

%% file: sections/7_Top_level_experiments/7_2_System_Adoption.tex
\subsection{Adoption and Scalability}

\begin{figure}[h!]
    \centering
    \includegraphics[width=0.7\linewidth]{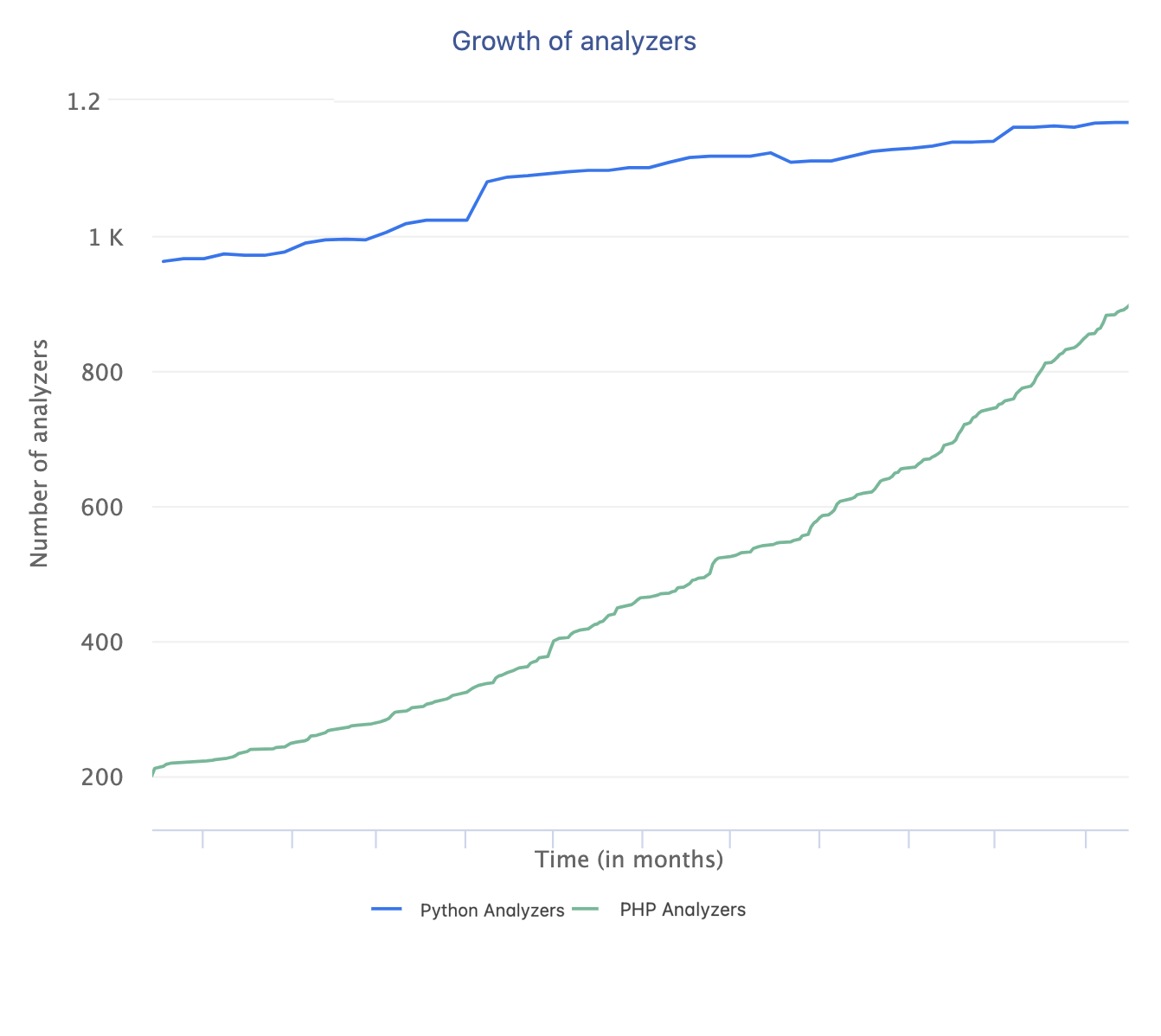}
    \caption{Analyzer Growth}
    \label{fig:analyzer_growth}
\end{figure}

DrP has been running in production for the past five years and user feedback and iterative development has helped improve many features. Figure \ref{fig:analyzer_growth} shows how this has helped steer growth in analyzers, for example a 2X growth seen in the past one year demonstrating success of the DrP platform. Overall, DrP has over 2000+ analyzers in production used by over 300 teams at Meta covering a diverse set of use cases.

Our measurements for 30 day moving average of the number of analyzer runs in our backend is 1.5 million which is roughly about 50K runs per day. These include over 250K alerts processed every 30 days. Our stand-alone CLI and UI is used by over 450 unique users per week, with UI users creating over 1.5K unique user sessions per week. All of these measurements show a healthy user engagement and usage of DrP.

Table \ref{tab:availablity} shows over 99.9\% availability of the DrP backend with most errors due to dropped requests attributed to loadshedding. 



Figure \ref{fig:overhead-slo} shows the runtime overhead for DrP backend executors in the worker tiers. Request queueing and binary execution can create bottlenecks but we have made several optimizations reducing overhead to acceptable levels.

\begin{table}[h]
    \centering
    \resizebox{\columnwidth}{!}{
        \begin{tabular}{|l|l|}
            \hline
            System Metrics & Value \\
            \hline
            Number of analyzers & 2000+ \\
            \hline
            Number of teams & 300+ \\
            \hline
            Number of analyzer runs & 1.5 million every 30 days \\
            \hline
            Number of alerts processed & 250K alerts every 30 days \\
            \hline
            Number of stand-alone CLI and UI users & 450 unique users per week \\
            \hline
        \end{tabular}
    }
    \caption{Summary of system metrics}
    \label{tab:metrics}
\end{table}

\begin{table}[h]
    \centering
    \begin{tabular}{|c|c|c|c|}
        \hline
        Endpoint & Success & Errors & Availability \\
        \hline
        Async & 1,464,832 & 228 & 99.98\% \\
        \hline
        Sync & 54,180 & 11 & 99.97\% \\
        \hline
    \end{tabular}
    \caption{Availability breakdown of service endpoints (for a 30 day period)}
    \label{tab:availablity}
\end{table}

%% file: sections/7_Top_level_experiments/7_1_System_Performance.tex
\subsection{MTTR}

\begin{figure}[h!]
    \centering
    \includegraphics[width=0.6\linewidth]{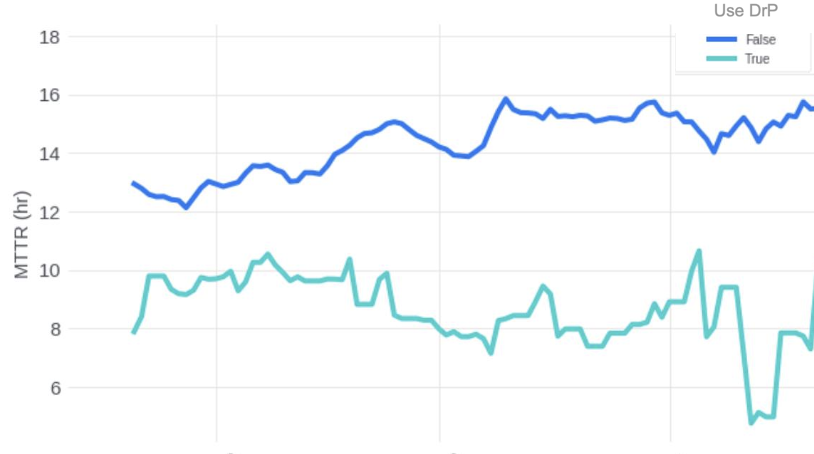}
    \caption{Teams using DrP see decline in average MTTR}
    \label{fig:mttr}
\end{figure}

MTTR stands for Mean Time To Resolution, a metric used to measure the average time it takes to resolve incidents. DrP automates manual investigations to triage incidents and thus helps to reduce MTTR. In this section, we demonstrate these benefits through a large-scale organizational study and a team-level breakdown analysis.

\subsubsection{Enhanced Experimental Design}

Our study employed a quasi-experimental design comparing MTTR of incidents using DrP versus those using traditional manual and ad-hoc investigation tools. This design is methodologically sound for several reasons:
\begin{itemize}
    \item \textbf{Real-world setting:} The study was conducted in a production environment with thousands of engineers and incidents over a full year, with consistent results replicated across multiple years.
    \item \textbf{Large sample size:} Our dataset provides robust statistical power with thousands of incidents, significantly larger than typical incident studies.
    \item \textbf{Control group:} The non-DrP group includes diverse teams across size, function, and on-call expertise, representing broader organizational practices while controlling for team-specific factors.
    \item \textbf{Consistent time frame:} Both groups were observed over identical time periods, controlling for seasonal variations and organizational changes.
    \item \textbf{Consistent incident complexity:} Only incidents requiring post-mortem reviews were analyzed, maintaining comparable severity levels across groups.
\end{itemize}

For our evaluations, we only consider incidents that went through a post-mortem process. The post-mortem process requires incidents to be updated with detection and resolution timestamps along with metadata such as processes and tools used for the resolution. MTTR is calculated as the time difference between detection and resolution timestamps. The recording of the timestamps and metadata is reviewed by post-mortem committees, ensuring high confidence in the accuracy of those numbers.

\subsubsection{Global Organization Study}
\label{mttr_large_scale_study}

In our first evaluation, we collected data for all incidents that occurred in at Meta for over a year, spanning thousands of teams and diverse use cases (services, AI/ML, data, mobile backends). We filtered out incidents with incomplete information, resulting in a high-quality dataset. Incidents were divided into two groups: those that used DrP during investigation and those that did not (using manual/ad-hoc tools). We calculated MTTR for each group by aggregating the mean time difference between resolution and detection timestamps.

Figure~\ref{fig:mttr} shows the results over time. MTTR for incidents using DrP is consistently and significantly lower than those not using DrP. Overall, the average MTTR improvement over the period of study was 20\%.

\subsubsection{Adoption-Level Impact Analysis}
\label{adoption_level_impact}

While our Meta-wide study demonstrates a 20\% average MTTR improvement, deeper analysis reveals that \textbf{adoption comprehensiveness significantly amplifies benefits}. Teams that onboarded DrP more comprehensively, with at least 10 analyzers, achieved substantially higher MTTR reductions ranging from \textbf{50\% to 80\%}. This finding suggests that DrP effectiveness is closely tied to adoption depth rather than mere usage.

Table~\ref{tab:team_mttr_study} shows the correlation between analyzer count and MTTR improvement. Teams with fewer than 5 analyzers showed modest improvements (10-15\%), while teams with 10+ analyzers consistently achieved transformative reductions. This pattern indicates that \textbf{comprehensive automation of investigation workflows}, rather than partial adoption, unlocks DrP's full potential.

\subsubsection{Team Breakdown Study}

To further validate our findings and account for team-specific factors, we conducted longitudinal analysis focusing on teams that adopted DrP. We identified \textbf{onboarding midpoints} for teams transitioning to DrP use, verified through interaction data and analyzer deployment timelines. For each team, we compared MTTR in equivalent periods before and after DrP adoption.

This approach addresses potential confounding variables by using \textbf{teams as their own controls}, eliminating systematic differences in expertise, problem complexity, and operational context. 

The longitudinal design strengthens our causal inference by controlling for time-invariant team characteristics while capturing the treatment effect of DrP adoption.

\begin{table}[h]
  \centering
  \resizebox{\columnwidth}{!}{%
    \small
    \begin{tabular}{|c|c|c|c|c|}
      \hline
      Team & MTTR Before (h) & MTTR After (h) & MTTR Improvement (\%) & Number of Analyzers \\
      \hline
      Team 1 & 771.63 & 139.22 & 81.98 & 136 \\
      \hline
      Team 2 & 613.78 & 154.19 & 74.87 & 92 \\
      \hline
      Team 3 & 621.91 & 97.138 & 84.31 & 66 \\
      \hline
      Team 4 & 488 & 154 & 68.5 & 48 \\
      \hline
      Team 5 & 329 & 139 & 57.8 & 39 \\
      \hline
      Team 6 & 107.95 & 29.06 & 73.08 & 29 \\
      \hline
      Team 7 & 44.32 & 28.4 & 56.1 & 23 \\
      \hline
      Team 8 & 286.64 & 267.34 & 7.35 & 12 \\
      \hline
    \end{tabular}
  }
  \caption{MTTR Reduction before and after using DrP.}
  \label{tab:team_mttr_study}
\end{table}

\subsubsection{Methodological Limitations and Validity Discussion}

\textbf{Threats to Validity:} While our quasi-experimental design cannot eliminate all potential confounds present in randomized controlled trials, several factors support the validity of our findings. \textbf{Selection bias} is mitigated by analyzing all incidents meeting post-mortem criteria rather than cherry-picked cases. \textbf{Temporal effects} are controlled through consistent observation periods and multi-year replication. \textbf{Measurement validity} is ensured through organizationally-reviewed timestamp data with high confidence levels.

The \textbf{control group validity} is supported by Meta's size and diversity while DrP serves 300+ teams, Meta encompasses significantly more teams, providing a substantial control population with varied expertise levels and operational practices.


%% file: sections/7_Top_level_experiments/7_4_Developer_Experience.tex
\subsection{Authoring Experience}

In this section, we provide some quantitative and qualitative measurements of the developer experience while authoring analyzers. DrP provides users with SDK, tools, and documentation to help develop analyzers quickly and accurately. Overall, we have found that users usually develop simple to medium analyzers within a day, and more complex analyzers in few days. The typical development cycle for teams is to first develop a basic analyzer and then iteratively improve it over time. It generally takes a few months for the analyzers to be fully evolved to handle all of the investigation workflows of the team. Providing boilerplate templates for the analyzers has helped a lot in quickly bootstrapping analyzers. Automated backtesting has also helped uncover bugs to improve the reliability and quality of analyzers.

\begin{figure}[h!]
    \centering
    \includegraphics[width=0.5\linewidth]{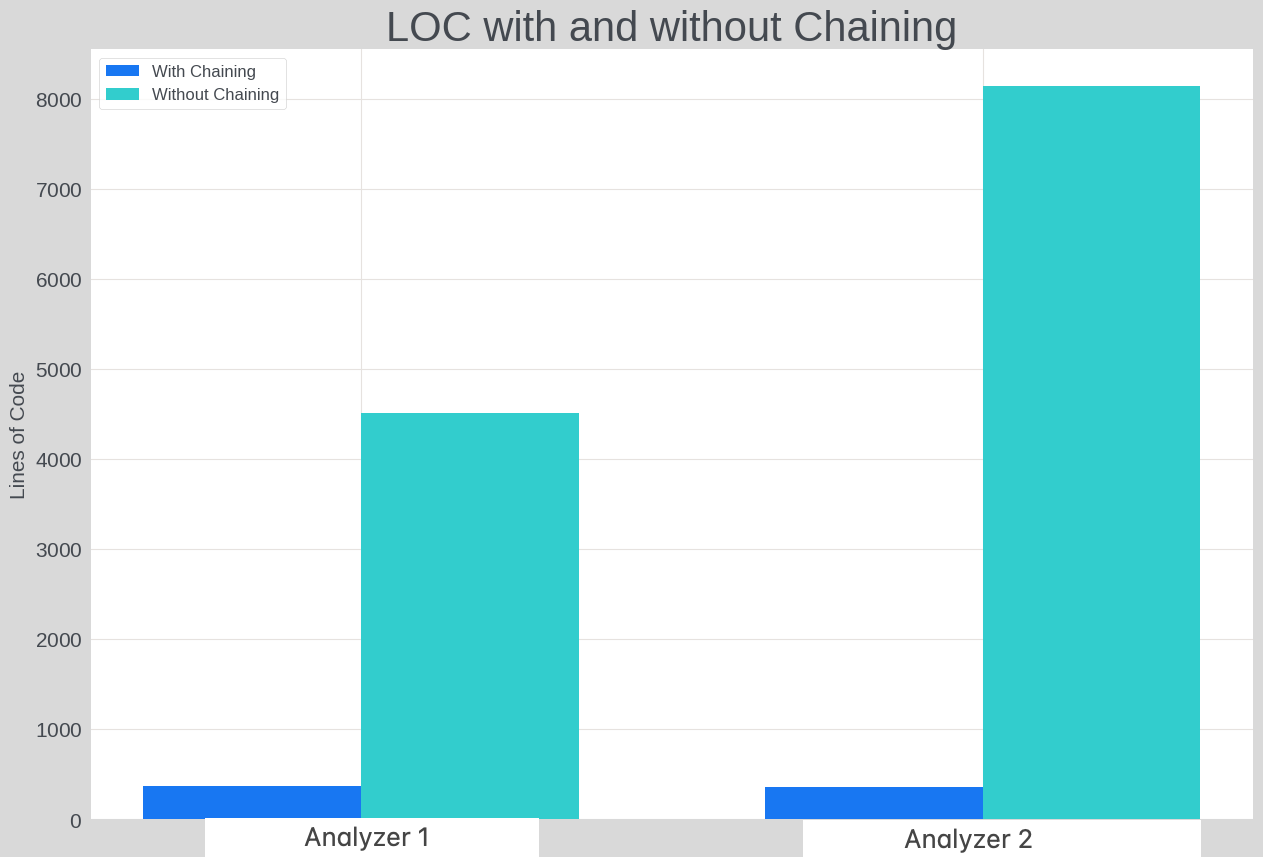}
    \caption{LOC before and after DrP SDK}
    \label{fig:loc}
 \end{figure}

One key feature helping developer experience has been re-usability provided by analyzer chaining. This has allowed quicker development by allowing analyzer authors to compose already available analyzers into their top level analyzer. In addition to code re-use, this improves maintainability and reduces code-complexity. Using lines of code (LOC) as a measure of complexity and code re-use, Figure \ref{fig:loc} shows how LOC improves with analyzer chaining vs without - 3 times improvement for 2 typical service debugging use cases. We have seen similar benefits across many analyzers with some of our power-users reporting 5-10x faster developer experience attributed to analyzer chaining after migrating their adhoc solution to DrP.
More than 21 percent of analyzers using analyzer chaining. 

\begin{figure}
    \centering
    \includegraphics[width=0.7\linewidth]{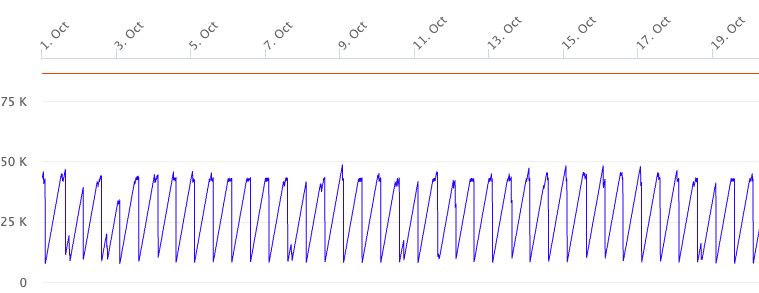}
    \caption{p99 Code Deployment time of ~12hrs}
    \label{fig:deployment-slo}
 \end{figure}

Another important aspect of developer experience is faster time from code merge to production. Because of the complex builds and canary testing built into our CI/CD system, deployment of analyzers can take time. In the early days of DrP, we had inefficiencies and failures in the deployment system and the long deployment times quickly became a key user dissatisfaction, sometimes taking days for code to reach production. With several improvements, the deployment time was reduced and we continuously monitor and measure this. Currently, our p99 deployment time is 12 hours (fig \ref{fig:deployment-slo}). 

\begin{figure}
    \centering
    \includegraphics[width=0.7\linewidth]{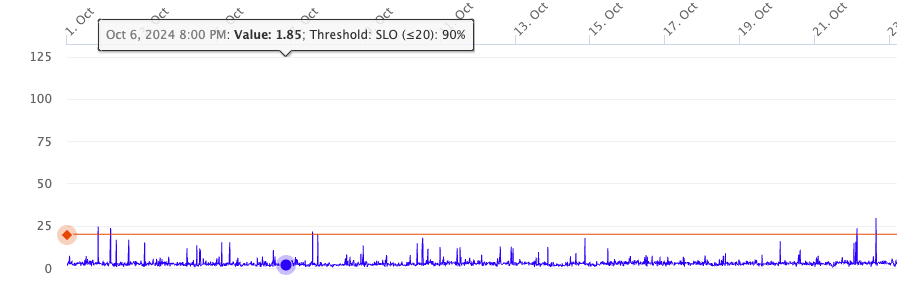}
    \caption{p99 Execution Overhead latency of ~20s}
    \label{fig:overhead-slo}
 \end{figure}

Another measurement we monitor is the overhead the underlying platform adds while running the analyzers. Developers care about runtime latency of their analyzers and we minimize any overheads caused by DrP. Currenly, our p99 overhead latency is about 20 seconds (fig \ref{fig:overhead-slo}). 

\begin{figure}
    \centering
    \includegraphics[width=0.5\linewidth]{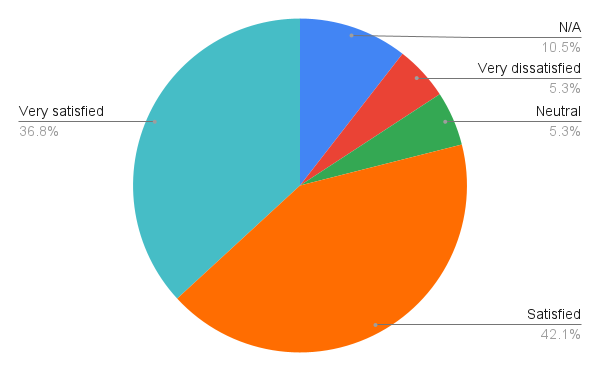}
    \caption{DrP Authoring User Satisfaction}
    \label{fig:user-authoring}
\end{figure}

We also regularly conduct developer experience surveys among users. We have consistently seen overall developer satisfaction numbers greater than 4.0/5.0. Figure \ref{fig:user-authoring} shows a breakdown of user satisfaction survey score for DrP authoring which shows roughly 80 percent users satisfied with the experience. Anecdotally, some of our power users have consistently given us positive feedback giving us high confidence in our framework and approach.

\subsection{On-call Experience}

DrP reduces the on-call effort during investigations saving engineering hours and reducing on-call fatigue. To illustrate this, we conducted a simulated experiment taking three investigation scenarios representing simple, medium, and complex analysis. (i) Simple - investigate service errors in a given time window. (ii) Medium - investigate container failures in shared pools (iii) Complex - investigate feature issues for AI/ML models

\begin{figure}[h!]
    \centering
    \includegraphics[width=0.5\linewidth]{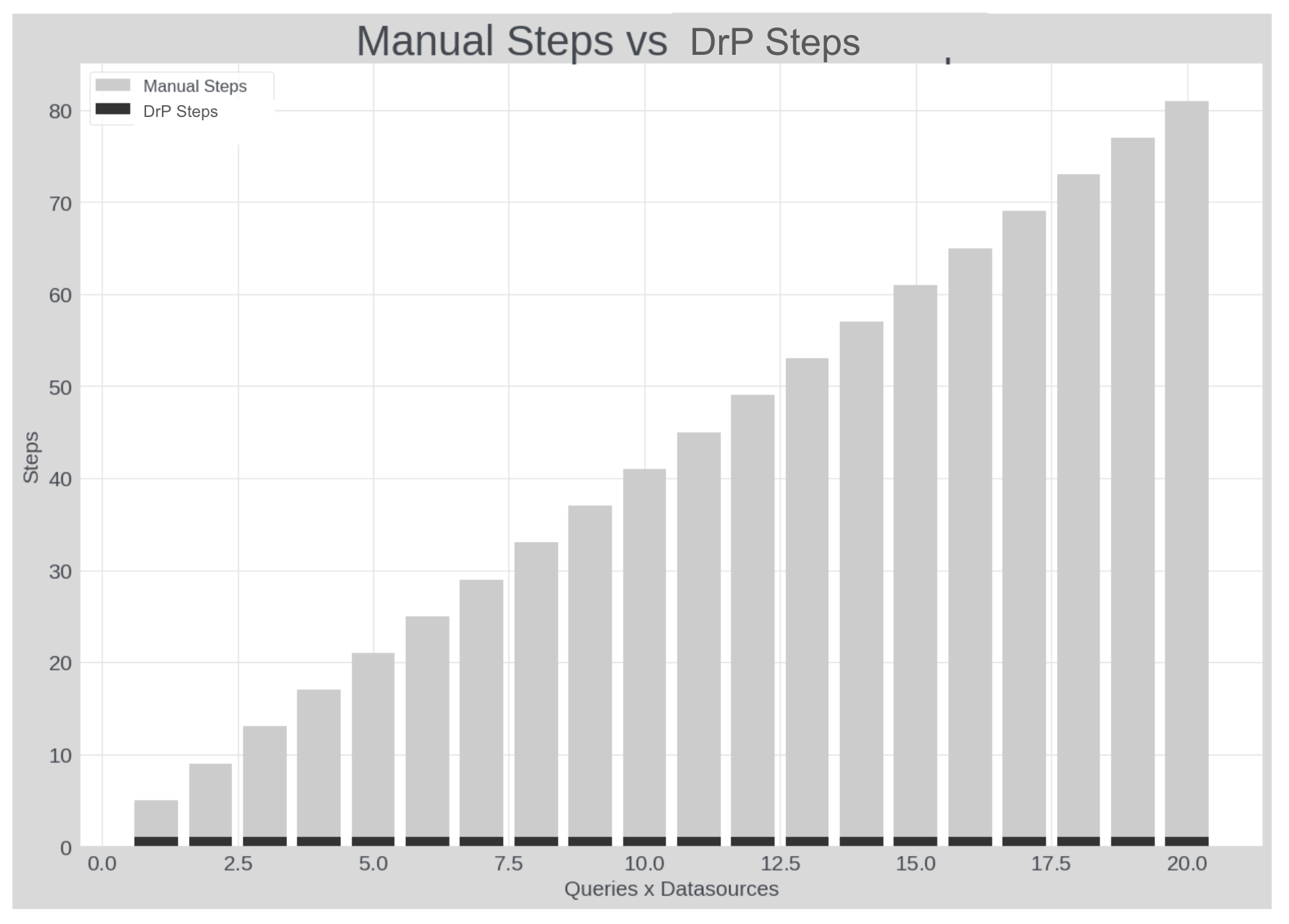}
    \caption{Number of steps for investigations - manual vs with DrP automation}
    \label{fig:steps-complexity}
\end{figure}

For each of these scenarios, we measured the number of steps (as a proxy for on-call load) taken by a manual approach (using the recommended playbook for that scenario) and compared it with using DrP instead. Figure \ref{fig:steps-complexity} illustrates the savings by DrP. The manual approach requires a large number of steps - preparing query filters, doing repeated queries, and repeated manual correlation. DrP on the other hand automates these steps and uses the underlying SDK's data access and analysis libraries for querying multiple data soures, doing dimensional analysis, and for timeseries/event correlation. This reduces the number of steps that the on-call engineer needs to do. In fact, in the scenarios above, with the DrP approach, the on-call engineer only needs to do one step - consume the automated results from the alert page. Figure \ref{fig:steps-complexity} also shows that as the complexity of investigations increase, the gains made by DrP in reduction of on-call load increases from 4X to 20X. If we compound this with the number of incidents that on-calls handle, the savings increase drastically. Also, to be noted is that DrP caters to a wide variety of use cases and thus the benefits seen from the framework span multiple on-call rotations and use cases.

 \begin{figure}[h!]
    \centering
    \includegraphics[width=0.5\linewidth]{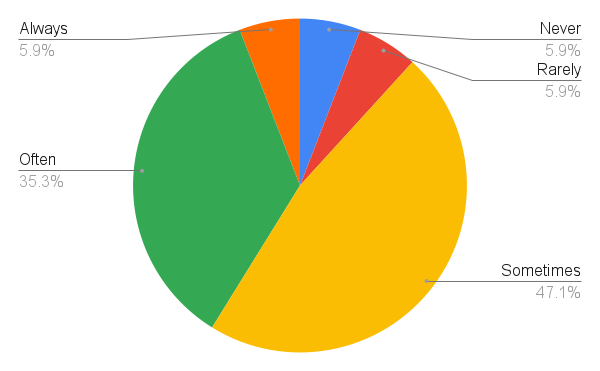}
    \caption{How often DrP is able to reduce MTTR during investigations}
    \label{fig:mttr-freq}
\end{figure}

In addition, similar to developer experience, regular surveys on on-call experience have been positive with overall satisfaction scores exceeding 4.0/5. One additional benefit that we have seen DrP provide is reduction in load on experts and removing them as bottlenecks during the investigation process. This allows experts to spend their engineering hours on other projects and allows less experienced and even junior engineers to handle complex investigations. This overall has also led to productivity and better engineering efficiency for many teams as shown in Figure \ref{fig:mttr-freq} where more than 80 percent developers report reduction in MTTR.

%% file: sections/7_1_User_Studies.tex
\label{case_studies}

In this section, we present a few case studies of DrP usage in production.

\par \noindent {\bf Platform A - Diagnosis Automation:} Platform A is used to manage campaigns. The Platform A Health team focuses on avoiding disruptions to the user experience. DrP helped reduce time to root cause by providing a guided flow to automate signal verification and diagnosis insights from hours to minutes for 92 percent of Platform A service events.

\par \noindent {\bf Service B - Reliability Improvements:} Service B provides external user facing services. DrP has significantly enhanced the reliability of Service B by automating outage investigation and recommending reroute of traffic to alternative services in under a minute. This automation saves hours of employee effort and improves user experience. By preventing 25,000 failed requests monthly, it has improved reliability and enhanced Service B's performance.

\par \noindent {\bf Platform C - ML Robustness:} Platform C used DrP to build a framework and drill-down components to help pinpoint the misbehaving ML prediction model based on calibration metrics. With DrP, the Platform C team was able to reduce the P50 time to isolate the problematic prediction models from 3 hours to 30 minutes.

\par \noindent {\bf Tool D - Diagnostics: } Tool D serves as a tool for managing Machine Learning (ML) workflow runs, experiments, models, and features, significantly impacting the ML development lifecycle. Tool D utilized DrP to aid in complex ML investigations during training, such as specific debugging tasks. The DrP team launched Tool D Diagnostics, from which approximately 2,000 automated analyses are triggered daily, powered by DrP, helping reduce mean time to resolve ML investigations.


%% file: sections/8_Lessons_learned.tex
\par \noindent {\bf Community based approach:} DrP's success has largely been due to its community-driven approach to analyzer development. Early on, the DrP team took the lead and bootstrapped adoption by building analyzers for common investigations, eg. for services infrastructure and hardware. However, to scale for diverse use cases, we developed the SDK that allowed teams to build custom analyzers and chain them together, leveraging community and expertise. The engaged effort by the community and democratization of analyzer development has led to success of DrP.

\par \noindent {\bf Meeting users where they are:} A key learning from DrP's adoption was the need to focus on developer experience. Unlike core product development, analyzer development is not the mainstream project for most software engineers. This has required that we focus on speed, simplicity, and flexibility for analyzer development. Additionally, multi-language SDK is essential providing users choice of language and prevent steep learning curves.

\par \noindent {\bf Integration into mainstream workflows:} Another reason for DrP's success has been integration into the mainstream operations workflows. This ensured that DrP got used where it mattered most and helped prioritize feature improvements. Early integration with alerting systems proved particularly valuable, providing immediate relief to on-call engineers and guiding design improvements to address painpoints.

\par \noindent {\bf Assist versus full automation:} Our initial vision was {\em complete automation} of the investigation workflows. During our journey, we learnt some practical lessons. First, with evolving systems, it is difficult to keep analyzers up-to-date. Second, statistical and machine learning analysis have limitations and do result in false positives and negatives. Third, culturally, engineers and on-calls don't always trust a fully automated system and some users prefer manual control or native tools for investigations. All of these have led us to pivot towards an assistive approach, where DrP offers insights and recommendations which are validated by engineers, balancing automation with human expertise.

\par \noindent {\bf Data is everything:} A key to seamless investigation is quality of data and metadata in telemetry systems. This is even more important when automating investigations. In our experience, we have faced several issues around observability data that has made it hard to correlate data from different sources. Similarly lack of structured metadata, e.g., for service dependencies or data lineage limits our ability to make downstream or upstream correlations.

\par \noindent {\bf Analyzer quality and maintenance:} Ensuring analyzer quality and maintenance is another crucial aspect necessary for DrP's success. To maintain analysis quality, we have used continuous feedback mechanisms from on-calls, and canary testing during deployment. Keeping analyzers up-to-date has had challenges that has needed continuous investment from teams. Like any other software development, teams need to make provisions for long term development and maintenance of the analyzer code. Also, teams need to estimate the right time to invest in analyzers. If the investigations are too simple and repetitive, a dashboard approach may be more effective, however as investigation complexity and team sizes grow, analyzers would add more value.

\par \noindent {\bf Do not over-index on AI based systems for diagnosis:} DrP supports several ML based analysis libraries in the SDK. Limitation of purely ML based systems is the quality of data being trained on as well as quality and structure of data available at investigation time. Another limitation is being able to customize workflows. Instead, we learned that building analyzers with a combination of rule based suggestions based on community expertise augmented with AI is a better approach. Supplementing the analysis with dashboards for e2e visualization has also proved effective.

%% file: sections/9_Related_Work.tex
There are several industry products in the operations space\cite{pagerduty, spunk, new-relic, datadog_watchdog, dynatrace, honeycomb, shoreline}. These products provide integrations for monitoring, alerting, and runbooks. DrP stands out as an automated investigation platform which enables authoring and orchestration of complex data analysis to investigate a wide range of system incidents. In comparison, the majority of industry tooling either focuses on workflow automation for repetitive tasks (restarting services, creating environments, support requests) or provide dashboards for generic data visualization which focuses more on observability of the system health vs. investigations. There are some chat based copilots for investigation built on top of data visualization solutions, which do a good job of fetching observability data through text based queries, but fall short of complex data analysis and automated investigations. In other products offering investigation capabilities, DrP provides distinct advantages through an expressive and flexible SDK, and proven deployment at large scale for a wide range of domains.
    
Cloud monitoring solutions\cite{aws-cloudwatch, gcp-monitoring, azure-monitoring} and systems such as Netflix Winston/Bolt\cite{netflix-winston} are also deployed at large scale, however, they do not provide as structured an approach to investigations as DrP does with SDK and integrations. DrP is on par in handling large data sets and data sources, and supporting ML workflows, along with other advantages.

In academia, automated anomaly detection and troubleshooting has been a subject of extensive research\cite{chandola_alerting, chengwei-troubleshooting}. Various techniques have been proposed to automate the process, such as detecting anomalies from the real-time telemetry data using as statistical or machine learning models\cite{hagemann2020systematic}\cite{soldani2022anomaly}\cite{toledano2018real}, correlating events with time series\cite{luo2014correlating}\cite{marvasti2013anomaly}, or configuration change analysis\cite{wang2004automatic,hu2020automated,mehta2020rex}. DrP can leverage any of these techniques as part of the SDK libraries. As an example, DrP SDK currently supports a few home-grown analysis libraries that have been catered to and tuned for the production system. These libraries can be enhanced or replaced by any of the existing related work as well.

Recently, large language models (LLMs) have demonstrated high generalizability in problem-solving across multiple data modalities and have been applied to assist on-call engineers in root-causing and mitigating production incidents \cite{ahmed2023recommending}\cite{roy2024exploring}\cite{googleMonitorTroubleshoot}. RCACopilot \cite{chen2024automatic} introduces a system that automatically matches incoming incidents with the predefined incident handlers based on the alert type to collect multi-modal diagnostic data to the incident, and utilize LLMs to predict a root cause category. DrP's analyzers are similar to incident handlers, however, we provide a more generalized framework to develop such analyzers and execute at scale. We are also actively looking into integrating LLM based libraries into the DrP SDK as part of ongoing work.

Several studies have proposed domain-specific deep dive root cause diagnosis systems, for example, using tracing and logs. Fay\cite{erlingsson2012fay} provides flexible and declarative tracing framework to monitor and debug distributed systems. ExChain\cite{li2024exchain} focused on a logging framework for diagnosing exception-dependent failures in online services. In big data platform domain, various specialized diagnosis system has been developed to enhance critical path analysis for streaming computations\cite{hoffmann2018snailtrail}, diagnosis job slowdowns using regression-based approach\cite{shao2019griffon}, and tackle computation skew \cite{teoh2019perfdebug}. Compared to these works, DrP provides a more generic abstraction at high level, enabling domain-specific experts to easily plug-in their specialized root cause analysis solution, and benefit from the end-to-end integration with alerting and management tools. In a sense, DrP' offer is complementary to these specialized systems. If specialized diagnosis systems provide analysis libraries for initial triage/mitigation, they can be integrated as part of the DrP SDK libraries.

%% file: sections/10_Conclusion.tex
We have described DrP, an end-to-end system for automating investigations at large scale. We discuss our novel solution consisting of an expressive and flexible SDK for authoring analyzers, a scalable backend system to execute the analyzers, integrations into mainstream workflows such as alerts and incident management tools, and a post-processing system to take actions on investigations. DrP has been deployed to serve diverse investigation use-cases across different domains and a large number of teams. DrP has been in production for the past 5 years and has helped drive impact by reducing MTTR for incidents, and has received positive feedback on reducing on-call toil. 

Looking ahead, we are continuing to evolve DrP and are extending it with Generative AI–based analysis to enable more powerful investigation automation and significantly reduce user onboarding costs.

%% file: sections/Acknowledgements.tex
We would like to thank all those who contributed to DrP systems including Akash Jothi, Kshitiz Bhattarai, Alex He, Juan-Pablo E, Oswaldo R,   Vamsi Kunchaparthi,  Daniel An, Rakesh Vanga, Ankit Agarwal, Narayanan Sankaran, Vlad Tsvang, Khushbu Thakur, Rohit JV, Bao Nguyen, Viraaj Navelkar, Arturo Lira, Nikolay Laptev, Sean Lee, Yulin Chen. We would also like
to thank the support of  John Ehrhardt, Ruben Badaró, Victoria Dudin, Gautam Shanbhag, Barak Yagour, Aparna Ramani. We are also indebted to CQ Tang, Mike Chow, Partha Kanuparthy, and anonymous reviewers for reviewing the paper and insightful comments.